\documentclass[preprint]{elsarticle}

\usepackage{hyperref}
\usepackage{amsmath}
\usepackage{color}

\begin{document}
	
	\begin{frontmatter}
		
		\title{Probing the scale of non-commutativity of space
			\vskip-1.8cm\hfill\small TIFR/TH/18-13\vskip1.5cm}
		

		\author[mymainaddress]{Pulkit S. Ghoderao}
		\ead{pulkitsg@iitb.ac.in}
		
		\author[mysecondaryaddress]{Rajiv V. Gavai}
		\ead{gavai@tifr.res.in}
		
		\author[mymainaddress]{P. Ramadevi}
		\ead{ramadevi@phy.iitb.ac.in}

		\address[mymainaddress]{Department of Physics, Indian Institute of Technology Bombay, Powai, Mumbai - 400 076}
		\address[mysecondaryaddress]{Department of Theoretical Physics, Tata Institute of Fundamental Research, Colaba, Mumbai - 400 005}
		
		\begin{abstract}
			Examining quantum electrodynamics in non-commutative (NC) spaces along with
			composite operators in these spaces, we show that i) any charge g for a
			fermion matter field is allowed provided the basic NC photon-photon
			coupling is g, however no other multiples of g are permitted and ii) composite 
			operators do not have a simple transformation which can be attributed to the 
			effective total charge of the composite particle. Taken together these results 
			place a limit on the scale of non-commutativity to be at most smaller that 
			current LHC limits for compositeness. Furthermore, they
			also suggest that a substructure at still smaller scales is needed if such
			spaces are to be a physical reality.
		\end{abstract}
		
		\begin{keyword}
			Non-commutative QED, Scale of non-commutativity, Quark-compositeness
		\end{keyword}
		
	\end{frontmatter}
	
	
	\section{Introduction}
	Non-commutative (NC) spaces arise naturally in the context of String Theory \cite{Seiberg:1999vs}. A useful method of studying such spaces is to capture the NC nature through a modification of commutation relations between the space-time coordinates, 
	\begin{align}
	\nonumber [x_\mu, x_\nu] = \iota \theta_{\mu\nu}.
	\end{align}
	Here $\theta_{\mu\nu}$ is a constant anti-symmetric matrix with real
	entries. By analogy with the position-momenta commutator relations in quantum
	mechanics, the scale of non-commutativity is expected to be captured by the
	$\theta$ element similar to $\hbar$ capturing the quantum scale. To see this
	more clearly, we separate $\theta_{\mu\nu} = \theta f_{\mu\nu}$, where
	$f_{\mu\nu}$ is a dimensionless constant antisymmetric matrix with matrix
	elements of ${\cal O}(1)$. $\theta$ 
	is a dimensionful  parameter determining the scale of NC space and is known as the \textit{non-commutative parameter}. This
	scale has the dimensions of area, $[L^2]$. A natural question that arises is
	whether observations from accelerator or spectroscopic experiments place any 
	constraint on the scale of non-commutativity. It will help formulate NC signatures in current experiments such
	as the Large Hadron Collider at CERN. \\
	
	It has been argued \cite{PhysRevLett.86.2716} using NC version of quantum
	electrodynamics, that Lamb Shift measurements in hydrogen atom can potentially
	reveal the NC scale. This result has been contested \cite{PhysRevLett.88.151602}
	where it was stated that particles with opposite charge have opposite sign of
	the non-commutative parameter $\theta$. Thus non-commutative effects cancel at
	commutator level itself when relative coordinates between the oppositely charged
	proton and electron are considered. On the other hand, proton is known to be
	composed of quarks with charges $+2/3$ and $-1/3$. Following the arguments of
	ref.\cite{Chaichian2004} one expects that non-commutativity should play a role at a certain scale. To investigate this role, we examine in some detail the quantum electrodynamics and composite particles on NC space. \\
	
	In this letter, we find two results that gauge covariance in NC spaces demands. The first is obtained by extending the \textit{charge quantisation}
	problem in NC quantum electrodynamics to include any charge magnitude $g$ of the
	fermionic fields provided the basic photon coupling is the same value, $g$. The
	second result arises from a failure of any composite operator to gauge transform
	similar to a particle with corresponding aggregate charge. These results are
	then used to argue that substructure at scale below a particular limit is needed
	for NC spaces to be a physical reality. An estimate of this limit provides us
	with the scale at which non-commutativity may be expected to play a role.
	
	\section{Charge quantisation in NC quantum electrodynamics}
	Maxwell action in NC quantum electrodynamics is given by,
	\begin{align}
	\label{eqn:2}
	S = -\frac{1}{4} \int d^4 x~ \left(F_{\mu\nu} \star F^{\mu \nu}\right),
	\end{align}
	with
	\begin{align}
	\label{eqn:3}
	F_{\mu\nu} = \partial_\mu A_\nu - \partial_\nu A_\mu - \iota g [A_\mu, A_\nu]_{\star},
	\end{align}
	where $[~,~]_\star$ indicates the Moyal bracket associated with the Moyal (star) product,
	\begin{align}
	\label{eqn:4}
	(f \star h) (x)= f(x) \exp{\left(\frac{\iota\theta^{\mu\nu}}{2} \overleftarrow{\partial_\mu} \overrightarrow{\partial_\nu}\right)}h(x).
	\end{align}
	Here $g$ is the photon-photon coupling. In the remainder of this section we follow the treatment in ref.\cite{Hayakawa:1999zf}, which established charge quantisation in NC spaces. Note that we have an arbitrary, real valued $g$ as against $g=1$ in ref.\cite{Hayakawa:1999zf}. The case of {\it arbitrary} $g$ has been ignored or seems to have been overlooked there.
	
	To include fermionic part of the action, covariant derivative can be defined in
	two ways \cite{SheikhJabbari:2000vi} owing to the fact that product between fields is no longer commutative:
	\begin{align}
	\label{eqn:5}
	D_\mu \psi &= \partial_\mu \psi - \iota n (A_\mu \star \psi),\\
	\label{eqn:5.1}D_\mu \hat{\psi} &= \partial_\mu \hat{\psi} - \iota n (\hat{\psi}\star A_\mu). 
	\end{align}
	Here $n$ denotes the fermionic charge associated with these fields. The fields
	$\psi$ and $\hat{\psi}$ possess different U(1) gauge transformations but are
	charge conjugate in NCQED \cite{SheikhJabbari:2000vi}. We let $n$ to have any
	value as well. The need for gauge invariance will be seen to fix it.\\

	The Maxwell action \eqref{eqn:2} can be checked to be invariant under gauge transformation for $A_\mu$ field given by,
	\begin{align}
	\label{eqn:6}
	A'_\mu = U^g \star A_\mu \star U^{-g} + \frac{\iota}{g} (U^g \star \partial_\mu U^{-g}).
	\end{align}
	Note the appearance of the power $g$ in the transformation above, as a
	generalization from the case of $g=1$.
	We choose proper ``fundamental" representations of the NC U(1) group \cite{SheikhJabbari:2000vi} having transformations,
	\begin{align}
	\label{eqn:7}
	\psi' &= U^n \star \psi,\\
	\label{eqn:7.1}\hat{\psi}' &= \hat{\psi} \star U^{n}.
	\end{align}
	In the above, we have defined finite gauge transformations using the operators $U(\lambda)$ which take the form, 
	\begin{align}
	\label{eqn:8}
	U^a = \exp{\left( \iota a \lambda(x)\right)}_\star = 1 + \iota a \lambda + \frac{(\iota)^2 a^2}{2!} (\lambda \star \lambda) + ... 
	\end{align}
	where $\lambda$ is the infinitesimal gauge parameter, $a$ is an arbitrary real scalar.
	
	\subsection{Preserving covariance of the derivative}
	It is natural to seek a gauge-covariant derivative for $\psi$ field. In order 
	to preserve covariance under a gauge transformation, one must demand
	\begin{align}
	\label{eqn:9}
	D'_\mu \psi' = U^n \star D_\mu \psi.
	\end{align}
	This must also equal the definition \eqref{eqn:5}, 
	\begin{align}
	\label{eqn:10}
	D'_\mu \psi' = \partial_\mu \psi' - \iota n (A'_\mu \star \psi').
	\end{align}
	Solving for \eqref{eqn:9} in the infinitesimal case\footnote{Infinitesimal case involves cutting off the expansion of $U$ operator \eqref{eqn:8} at linear order in $\lambda$.} we obtain,
	\begin{align}
	\nonumber D'_\mu \psi' =  ~&\partial_\mu \psi - \iota n (A_\mu \star \psi) + \iota n (\lambda \star \partial_\mu \psi) \\
	&+ n^2 (\lambda \star A_\mu \star \psi).
	\end{align}
	On the other hand from \eqref{eqn:10}, in the infinitesimal case,
	\begin{align}
	\nonumber D'_\mu \psi' = ~&\partial_\mu \psi  - \iota n (A_\mu \star \psi) +  \iota n (\lambda \star \partial_\mu \psi)\\
	&+ n (n-g) (A_\mu \star \lambda \star \psi) + ng (\lambda \star A_\mu \star \psi).
	\end{align}
	To achieve equality between the last two expressions,
	\begin{align}
	n(n-g) = 0 \Rightarrow n=0 ~\text{or}~ n= +g.
	\end{align}
	Performing a similar exercise for the $\hat{\psi}$ field using \eqref{eqn:5.1} and \eqref{eqn:7.1}, we obtain $n=0 ~\text{or}~ n= -g$. Thus, $n = \pm g, 0$.\\

	This gives our first result, extending the established `\textit{charge quantisation}' in NC theories to include not only $\pm1$ as stated by
	ref.\cite{Hayakawa:1999zf} but also any possible charge value subject to the
	condition that it must exactly equal the photon-photon coupling.  
	
	\section{Composite operator in NC space}
	To resolve the issue surrounding the role fractionally charged quarks may
	possibly play in leading to nontrivial results due to non-commutativity for the
	Lamb shift of hydrogen atom, a basic requirement is the construction of a
	composite operator for the proton. As usual, we assume that the $SU(3)$ colour
	gauge theory confines quarks and proton in their bound state.  However, the
	quark fields may be defined on NC space, and lead to interesting consequences
	for Lamb shift. As in quantum chromodynamics (QCD), we
	construct the proton operator to be a composite of one down and two up quarks.
	The star product which is characteristically a point-wise multiplication allows
	us to define this operator at a particular point as,
	\begin{align}
	\label{eqn:11}
	\psi_p = \psi_u \star \tilde{\psi}_u \star \psi_d.
	\end{align}	
	Under a $U(1)$ gauge transformation we expect the proton to transform with 
	charge $+1$,
	\begin{align}
	\label{eqn:12}
	\psi'_p = U \star \psi_p.
	\end{align}
	This must also equal the definition in \eqref{eqn:11},
	\begin{align}
	\label{eqn:13}
	\psi'_p &= \psi'_u \star \tilde{\psi'}_u \star \psi'_d \\
	&= (U^{2/3} \star \psi_u) \star (U^{2/3} \star \tilde{\psi}_{u}) \star (U^{-1/3} \star \psi_d).
	\end{align} 
	Solving for \eqref{eqn:12} in the infinitesimal case,
	\begin{align}
	\label{eqn:14}
	\psi'_p &= \psi_u \star \tilde{\psi}_u \star \psi_d + \iota (\lambda \star \psi_u \star \tilde{\psi}_u \star \psi_d).
	\end{align}
	On the other hand from \eqref{eqn:13}, under infinitesimal case,
	\begin{align}
	\label{eqn:15}
	\nonumber \psi'_p =~& \psi_u \star \tilde{\psi}_u \star \psi_d - \frac{\iota}{3} (\psi_u \star \tilde{\psi}_u \star \lambda \star \psi_d) \\
	&+ \frac{2\iota}{3} (\psi_u \star \lambda + \lambda \star \psi_u) \star (\tilde{\psi}_u \star \psi_d).
	\end{align}
	It is immediately apparent that \eqref{eqn:14} and \eqref{eqn:15} cannot become
	equal. This problem persists even if we try to take an arbitrary linear
	combination of the $u$ and $d$ quarks in defining the composite \eqref{eqn:11}.
	The issue arises from the fact that ordering between gauge parameter and quark
	fields under the star product is different for the above two expressions. Being
	intimately linked with the non-associativity property of star product, we
	believe it is in principle not possible to get around this issue. This suggests
	that a proton composite is forbidden for quarks living in NC space.\\
	
	Extending this result, it also seems impossible to construct any composite operator as its gauge transform must necessarily involve different ordering between constituent fields and the gauge parameter. Hence existence of composite particles appears forbidden in NC space.
	
	\subsection{An approximate condition for composite particles to exist in NC space} 
	
	Consider a composite operator. As a case in point, let it be a proton composite. The star product between the gauge parameter and one of its constituent fields, $\lambda$ and $\psi$, is $\lambda \star \psi$.
	Using the definition of star product in \eqref{eqn:4} this becomes,
	\begin{align}
	\lambda \star \psi = \lambda \exp{\left(\frac{\iota\theta^{\mu\nu}}{2} \overleftarrow{\partial_\mu} \overrightarrow{\partial_\nu}\right)} \psi.
	\end{align}
	Expanding the exponential in terms of $\theta$,
	\begin{align}
	\label{eqn:16}
	\lambda \star \psi = \lambda \left(1 +  \frac{\iota\theta^{\mu\nu}}{2} \overleftarrow{\partial_\mu} \overrightarrow{\partial_\nu} + \mathcal{O}(\theta^2)\right) \psi,
	\end{align}
	similarly, a star product between $\psi$ and $\lambda$ is expanded as,
	\begin{align}
	\psi \star \lambda = \psi \left(1 +  \frac{\iota\theta^{\mu\nu}}{2} \overleftarrow{\partial_\mu} \overrightarrow{\partial_\nu} + \mathcal{O}(\theta^2)\right) \lambda.
	\end{align}
	To bring equality between \eqref{eqn:14} and \eqref{eqn:15}, the effects of ordering with respect to star product must become negligible. In other words, 
	\begin{align}
	\label{eqn:17}
	\lambda \star \psi - \psi \star \lambda =  \iota \theta^{\mu\nu} \partial_\mu \lambda~ \partial_\nu \psi + \mathcal{O}(\theta^3) \longrightarrow 0.
	\end{align}
	If the particle characterized by the composite operator, proton in this case,
	has a radius $r_p$, then by uncertainty principle one may expect the typical
	momenta of its constituents, which are equivalent to gradient of $\psi$, to be
	of $\mathcal{O}(r_p^{-1})$. Thus for \eqref{eqn:17} to be negligible,  
	\begin{align}
	\label{eqn:18}
	\theta << r_p^2.
	\end{align}
	Therefore, it is possible to get away from the issue of transformation of
	composite particles, albeit in an approximate way if the length scale of
	non-commutativity is much less than radius of the composite particle. Note that
	in such a case, one has effectively usual QED at the scale $ \sim r_p$, with a
	form factor for the bound state.  NCQED effects may only be seen through deep
	inelastic scattering which probes the underlying quark structure.  
	
	\section{Scale of NC space}
	Armed with the above arguments, we can revisit the question of fixing a scale for the non-commutativity parameter $\theta$. We list the successive possibilities,  
	\begin{itemize}
		\item \textit{Hypothesis 1: NC scale is fixed at the nuclear scale itself.}
		\begin{itemize}
			\item Condition \eqref{eqn:18} is not satisfied for the proton and other hadrons. Hence we cannot treat them as composite particles. To develop theories using this hypothesis we need to consider proton and other hadrons like $\Delta^{++}$ to be point particles. 
			\item In this case, the coupling $g$ equals both proton and electron charge, $n = 1$. As shown in ref.\cite{PhysRevLett.88.151602}, this will imply that there is no correction to the Lamb shift.
			\item \textit{Caveat}: However this hypothesis fails to account for charge quantisation as there are particles like $\Delta^{++}$ that have a charge magnitude of $n=2$.
		\end{itemize}
		\item \textit{Hypothesis 2: NC scale is fixed well below the nuclear
			scale but at or above the compositeness scale \footnote{Compositeness scale of
				quarks and leptons means the length scale above which these particles are
				effectively point particles, i.e., without any substructure. This scale is
				approximately $10^{-20}m$ as we describe later.} of quarks and leptons.}
		
		\begin{itemize}
			\item In this case, \eqref{eqn:18} is satisfied for proton and other hadrons but not for quarks and leptons. Hence to develop theories using this hypothesis we are allowed to treat proton and other hadrons to be composite particles whereas quarks and leptons have to remain point-like.
			\item \textit{Caveat}: Even if we account for all the leptons using $n = g =1$, the charge quantisation problem becomes severe as quarks themselves have varying magnitude of fractional charges and these are different from the lepton charge magnitude of $n=1$.
		\end{itemize}	
		\item \textit{Hypothesis 3: NC scale is fixed well below the compositeness scale of quarks and leptons.}
		\begin{itemize}
			\item Condition \eqref{eqn:18} is satisfied for proton and other hadrons as well as for quarks and leptons. Thus, to develop theories using this hypothesis, it is allowed to treat quarks and leptons as composite particles along with proton and other hadrons.
			\item The charge quantisation problem can be resolved by choosing suitable substructure for the quarks and leptons. We shall see how this is achieved in the following section.	
		\end{itemize}
	\end{itemize}
	
	Various compositeness scales for quarks and leptons have been 
	probed by LHC experiments \cite{rpp18web} and suggest a range of its 
	energy scale $\Lambda$ between $10-25$ TeV, or equivalently, length scale between $7.9-19.7 \times 10^{-21} m$. With this information we can fix the NC scale to necessarily lie below $2 \times 10^{-20}m$.
	
	\section{Possible constructions for matter in NC space}
	\subsection{A possible preon construction}
	As mentioned in the previous section, it is possible to resolve the charge quantisation problem by suitably choosing substructure to quarks and leptons. In the following, we take a look at one such possible candidate substructure mainly to demonstrate the plausibility of our argument. In this model, the NC scale is fixed well below quark and lepton compositeness scale. Quarks and leptons can thus be treated as composite particles. The substructure of matter comprises entirely of two new ``fundamental" particles with charges $+1/6$ and $-1/6$. We note that the charge quantisation problem is solved owing to same charge magnitude assignment for both particles. The quarks are constructed by a combination of four while the leptons through a combination of six such particles. For example, a down quark comprises of three similar particles with charge $-1/6$ and one different particle with charge $+1/6$. In a similar fashion all composites and charge assignments in the Standard Model can be obtained.\\
	
	Alternatively if neutral particles are also used as constituents of quarks and leptons, then other preon models such as the elaborate Rishon model \cite{Harari:1981uh} could possibly describe matter in NC space. This model has ``fundamental" particles with charges $\pm 1/3$ and $0$, thus satisfying the charge quantisation in NC space.
	
	\subsection{Possible construction using Han-Nambu integer charged quarks}
	The Han-Nambu integer charged quarks are an alternative to the $SU(3)_c$ quarks of the Standard Model. This model \cite{PhysRev.139.B1006} consists of three triplets of quarks all having the charge magnitude of $1$. As is apparent, such a construction immediately solves the charge quantisation problem as all leptons too have charge magnitude of $1$. We note that it is not necessary for the Han-Nambu quarks to have substructure as charge quantisation problem is solved even if they are treated to be point-like. Hence, both hypotheses 2 and 3 of the previous section are plausible. The estimate of NC scale for such a construction can thus be modified to only lie well below nuclear scale.

	In conclusion, NC spaces may be sustainable either for preon models 
	where one may hope to explain the charge as well as other global quantum 
	numbers of both quarks and leptons suitably in terms of fields of charge $g 
	\le 1/3$ and $g = 0$ or for Han-Nambu type integrally charged quarks.

	\section{Conclusion and Discussion}
	In this letter, we have placed a limit on NC scale which can be improved in future at LHC or other high energy colliders as it is tied with compositeness of quarks and leptons. The discovery of substructure to quarks or electrons could be an indicator of possible advent of the NC regime. The current limit at which NC scale can be fixed is $2 \times 10^{-20}m$. Extensions of this work can involve constructing an explicit Model of particles which form the substructure of matter near NC scale.\\
	
	The QCD scale (like radii of low lying states) is typically 1 fm ($=
	10^{-15}m$), whereas we have placed an estimate of NC scale below $10^{-20}m$.
	We see that the non-commutative effects from the quark sector are unlikely to be
	pinned down unless computations in QCD reach a precision of 1 in $10^{10}$ or so
	since a composite operator would be unable to probe non-commutativity for lesser
	precision as shown in \eqref{eqn:18}. As a consequence, the claim in
	ref.\cite{Chaichian2004} stating QCD can reveal NC corrections to Lamb Shift appears
	difficult to achieve in practice at the current level of theoretical and
	experimental precision. We have also cast doubt on
	the possibility of no correction stated in ref.\cite{PhysRevLett.88.151602} as
	it fails to consider any composite nature of proton as well as quarks or
	leptons. Thus in order to study hydrogen spectrum in NC space, we need to explore new gauge theories at the NC scale as discussed in section 5.

	\section*{Acknowledgements}
	
	One of us (RVG) would like to thank Prof. Frithjof Karsch and other colleagues
	from the Fakult\"at f\"ur Physik of the Universit\"at Bielefeld for their kind
	hospitality where this manuscript was finished. This work was partially
	supported by the Sir J. C. Bose Fellowship of the Government of India. PR would like to thank Keshav Dasgupta of Mcgill University for their hospitality while finalising this manuscript.
	
	\section*{References}
	
	\bibliography{plb_letter_bib}
	
\end{document}